\documentclass{mn2e}
\input{epsf}
\usepackage{amssymb}

\newbox\grsign \setbox\grsign=\hbox{$>$} \newdimen\grdimen \grdimen=\ht\grsign
\newbox\simlessbox \newbox\simgreatbox
\setbox\simgreatbox=\hbox{\raise.5ex\hbox{$>$}\llap
     {\lower.5ex\hbox{$\sim$}}}\ht1=\grdimen\dp1=0pt
\setbox\simlessbox=\hbox{\raise.5ex\hbox{$<$}\llap
     {\lower.5ex\hbox{$\sim$}}}\ht2=\grdimen\dp2=0pt

\def\mathbi#1{\textbf{\em #1}}

\newcommand{\hMpc}{{\ifmmode{h^{-1}{\rm Mpc}}\else{$h^{-1}$Mpc }\fi}}
\newcommand{\hkpc}{{\ifmmode{h^{-1}{\rm kpc}}\else{$h^{-1}$kpc }\fi}}
\newcommand{\hMsun}{{\ifmmode{h^{-1}{\rm {M_{\odot}}}}\else{$h^{-1}{\rm{M_{\odot}}}$}\fi}}
\newcommand{\Msun}{{\ifmmode{{\rm {M_{\odot}}}}\else{${\rm{M_{\odot}}}$}\fi}}

\title[The mass and anisotropy profiles of galaxy clusters]
{The mass and anisotropy profiles of galaxy clusters from the projected phase space density: testing the method
on simulated data}
\author[R. Wojtak et al.]{Rados{\l}aw Wojtak,$^{1}$ Ewa L. {\L}okas,$^{1}$
Gary A. Mamon$^{2,3}$  and Stefan Gottl\"ober$^{4}$
\\   \\
$^1$Nicolaus Copernicus Astronomical Center, Bartycka 18, 00-716 Warsaw, Poland\\
$^2$Institut d'Astrophysique de Paris (UMR 7095: CNRS and Universit\'e Pierre \& Marie Curie),
    98 bis Bd Arago,
    F-75014 Paris, France \\
$^3$Astrophysics 7 BIPAC, University of Oxford, Keble Rd, OX1 3RH, Oxford, UK \\
$^4$Astrophysikalisches Institut Potsdam, An der Sternwarte 16, 14482 Potsdam, Germany\\
}

\begin{document}

\maketitle

\begin{abstract}
We present a new method of constraining the mass and velocity anisotropy profiles of galaxy clusters from
kinematic data. The method is based on a model of the phase space density
which allows the anisotropy to vary with radius between two asymptotic values. The characteristic scale
of transition between these asymptotes is fixed and tuned to a typical anisotropy profile resulting
from cosmological simulations. The model is parametrized by two values of anisotropy, at the centre of the
cluster and at infinity, and two
parameters of the NFW density profile, the scale radius and the scale mass.
In order to test the performance of the method in reconstructing the true cluster parameters we analyze
mock kinematic data for 20 relaxed galaxy clusters generated
from a cosmological simulation of the standard $\Lambda$CDM model. We use Bayesian methods of inference
and the analysis is carried out following
the Markov Chain Monte Carlo approach. The parameters of the mass profile are reproduced quite well, but we
note that the mass is typically underestimated by 15
percent, probably due to the presence of small velocity substructures. The constraints on the anisotropy profile for
a single cluster are in general barely conclusive. Although the central asymptotic value is determined
accurately, the outer one is subject to significant systematic errors caused by substructures
at large clustercentric distance. The anisotropy profile is much better constrained if one performs joint
analysis of at least a few clusters. In this case it is possible to reproduce
the radial variation of the anisotropy over two decades in radius inside the virial sphere.
\end{abstract}

\begin{keywords}
galaxies: clusters: general -- galaxies: kinematics and dynamics -- cosmology: dark matter
\end{keywords}

\section{Introduction}

Kinematic data play an important role in dynamical studies of galaxy clusters. They offer
a unique possibility to constrain simultaneously the total (dark and luminous) mass profile
and the orbital structure of galaxies which is commonly quantified by the anisotropy of velocity
dispersion tensor (Binney \& Tremaine 2008). The main challenge of this approach is the fact
that both factors, the mass and the anisotropy profile, are interconnected. In the particular
case of commonly used Jeans analysis of velocity dispersion profile this leads to a well known
problem of the mass-anisotropy degeneracy that hampers the constraining power of the data (e.g.
Binney \& Mamon 1982; Merritt 1987; Merrifield \& Kent 1990). It turns out that in general there are
two ways to break this degeneracy: one can either combine the results with other constraints on the
mass profile or use a dynamical model going beyond the Jeans equation.

The first solution has been widely applied in the literature. Promising constraints on the anisotropy
of galactic orbits in clusters have been obtained by combining the results from velocity dispersion
profiles with the mass constraints from X-ray gas (e.g. Benatov et al. 2006; Hwang \& Lee 2008) or lensing
data (e.g. Natarajan \& Kneib 1996). A different approach was adopted by Biviano \& Katgert (2004) who argued
for an isotropic velocity dispersion tensor of early type galaxies in clusters
from the ESO Nearby Abell Cluster Survey (ENACS). Assuming this property they inferred the total
mass profile from velocity dispersions of ellipticals and used this result to constrain the anisotropy
profile of late type galaxies. This approach was an implementation of the anisotropy inversion
algorithm (Binney \& Mamon 1982; Solanes \& Salvador-Sol\'e 1990).

The second method of breaking the mass-anisotropy degeneracy relies on the extension of the classical Jeans
formalism. The first natural step in this field is to consider the fourth velocity moment or kurtosis
(Merrifield \& Kent 1990). The method based on the joint fitting of velocity moments (dispersion and
kurtosis) was shown to provide constraints both on the mean anisotropy of a system and the parameters of the
mass profile ({\L}okas 2002; {\L}okas \& Mamon 2003; Sanchis, {\L}okas \& Mamon 2004; {\L}okas et al. 2006).
These results confirmed the idea that any attempt to infer the anisotropy profile from kinematic data of spherical
systems must be preceded by the construction of a detailed dynamical model. In general there are two ways to achieve
a desired complexity of a model. One is the so-called Schwarzschild modelling (Schwarzschild 1979) in which
one considers a superposition of base orbits defined in the integral space (e.g. Merritt \& Saha 1993;
Gerhard et al. 1998; Chanam\'e, Kleyna \& van der Marel 2008). Another one, which we adopt in this work, is to
provide a properly parametrized form for the phase space density.
%sufficiently general and anisotropic model of the phase space density $f({\mathbi r},{\mathbi v})$.

There were several studies devoted to the analysis of kinematic data in terms of the phase
space density. An important step in this field was made by Kent \& Gunn (1982) who used a family of simple
analytical models of the distribution function to analyze the data for the Coma cluster. Van der Marel et al.
(2000) obtained constraints on the anisotropy
of 16 galaxy clusters from the CNOC1 (Canadian Network for Observational Cosmology) cluster redshift
survey. A conceptually similar analysis was carried out by Mahdavi \& Geller (2004) for galaxy groups and clusters.
In all cases a constant anisotropy was assumed which does not reproduce well the results of
cosmological simulations where the dependence of the anisotropy on radius is usually seen (e.g. Mamon \& {\L}okas 2005;
Wojtak et al. 2005; Ascasibar \& Gottl\"ober 2008). Since the anisotropy profile has recently become a subject of
growing interest, it appears reasonable to generalize the above methods so that both the mass and
the anisotropy profiles may be inferred from the data. This implies that one has to deal with an anisotropic
model of the phase space density which accounts for its radial variation. Quite recently
several models satisfying this requirement have been proposed. The anisotropy profile is specified by
a proper parametrization of the angular momentum part of the distribution function (Wojtak et al. 2008)
or the augmented density (Van Hese, Baes \& Dejonghe 2009). The purpose of the present work was to adopt the approach of
Wojtak et al. (2008) to the Bayesian analysis of kinematic data and to test on mock data sets how well the mass
and anisotropy profiles are reproduced.

The paper is organized as follows. In the first section we introduce a model of the phase space
density and discuss its projection on to the plane of sky. In section 2 we describe mock kinematic data of
galaxy clusters generated from a cosmological simulation. Section 3 provides technical details
on the Monte Carlo Markov Chain analysis and section 4 presents the results. The discussion follows in
section 5.

\section{The phase space density}

\begin{figure}
\begin{center}
    \leavevmode
    \epsfxsize=8cm
    \epsfbox[50 50 580 610]{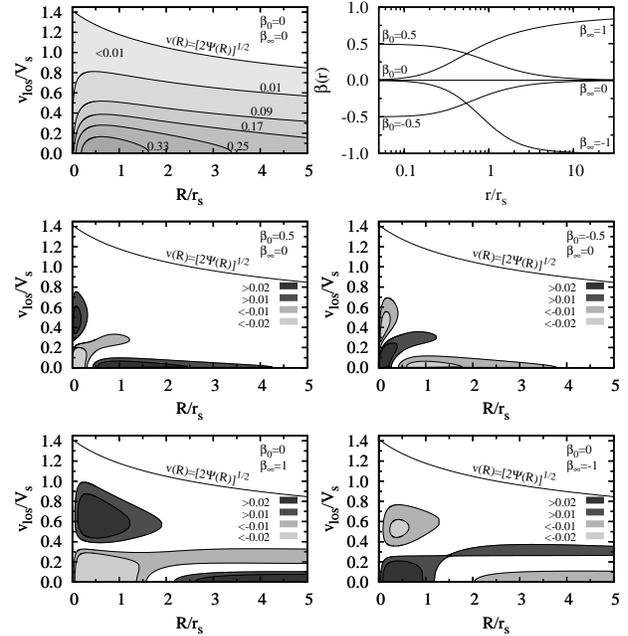}
\end{center}
\caption{The projected phase space density $f_{\rm los}(R,v_{\rm los})$ for five anisotropy profiles
characterized by different combinations of $\beta_{0}$ and $\beta_{\infty}$ and plotted in the top right
panel. The top left panel is a contour map of the isotropic $f_{\rm los}$ ($\beta_{0}=\beta_{\infty}=0$) and
the four bottom panels show the differences between a given $f_{\rm los}$ and the isotropic one.
The solid lines labelled by $v(R)=\sqrt{2\Psi(R)}$ are the profiles of the maximum escape velocity.}
\label{df_los_ex}
\end{figure}

Any spherical system in equilibrium embedded in a fixed gravitational potential
is described completely by the distribution function which depends on the phase space coordinates through
the binding energy $E$ and the absolute value of the angular momentum $L$. In this work, we use the model of
the distribution function recently proposed by Wojtak et al. (2008) which was shown to recover
spherically averaged phase space distribution of dark matter particles in simulated cluster-size haloes.
The main idea of this approach lies in the assumption that the distribution function is separable in energy
and angular momentum, i.e. $f(E,L)=f_{E}(E)f_{L}(L)$. The angular momentum part $f_{L}(L)$ is given by an
analytical ansatz motivated by the purpose of providing an appropriate parametrization of the
anisotropy profile, which is traditionally quantified by the so-called anisotropy parameter
\begin{equation}
\beta(r)=1-\frac{\sigma_{\theta}^{2}(r)}{\sigma_{r}^{2}(r)},
\end{equation}
where $\sigma_{\theta}$ and $\sigma_{r}$ are dispersions of the tangential and radial velocity respectively.
This part of the distribution function takes the following form
\begin{equation}	\label{f_L}
	f_{L}(L)=\Big(1+\frac{L^{2}}{2L_{0}^{2}}\Big)^{-\beta_{\infty}+\beta_{0}}
	L^{-2\beta_{0}},
\end{equation}
where $\beta_{0}$ and $\beta_{\infty}$ are the asymptotic values of the anisotropy parameter at
the halo centre and at infinity respectively. The scale of transition between these two
asymptotes is determined by $L_{0}$, whereas a typical radial range of the growth or decrease of $\beta(r)$ is
fixed at about 2 orders of magnitude centred on the radius corresponding to $L_{0}$ (see the anisotropy
profiles in the top right panel of Fig.~\ref{df_los_ex}). Although some recent models of the distribution
function offer a little more flexible parametrization of the anisotropy profile (e.g. Baes \& Van Hese 2007;
Van Hese et al. 2009), we find that our choice is quite suitable for the purpose of this work, given that we
wish to reproduce the variability of $\beta(r)$ with as few parameters as possible.

The energy part of the distribution function $f_{E}(E)$ is given by the solution of the integral
equation
\begin{equation}\label{rho_DF}
	\rho(r)=\int\!\!\!\int\!\!\!\int f_{E}(E)
	\Big(1+\frac{L^{2}}{2L_{0}^{2}}\Big)^{-\beta_{\infty}+\beta_{0}}
	L^{-2\beta_{0}}\textrm{d}^{3}v.
\end{equation}
This equation can be simplified to the one-dimensional integral equation and then solved numerically
for $f_{E}$ (see Appendix B in Wojtak et al. 2008). As an approximation for the density profile in (\ref{rho_DF}) we use
the NFW profile (Navarro, Frenk \& White 1997), i.e.
\begin{equation}\label{NFW}
	\rho(r/r_{s})=\frac{1}{4\pi(\ln2-1/2)} \frac{1}{(r/r_{s})(1+r/r_{s})^{2}}
	\frac{M_{\rm s}}{r_{\rm s}^{3}},
\end{equation}
where $r_{s}$ is the scale radius and $M_{s}$ is the mass enclosed in a sphere of this radius.
Both parameters of the mass profile provide natural scales of the phase space coordinates so that
any change of $M_{s}$ or $r_{s}$ corresponds to the expansion or contraction of a system in the
space of velocities or the positions. For the sake of convenience, we use the scaling which fixes
the range of positively defined binding energy per unit mass $E$ at $[0,1]$, namely
$r_{s}$ as the unit of radius and $V_{s}=(GM_{s}/r_{s})^{1/2}(\ln 2-1/2)^{-1/2}$ as the unit of
velocity (see Wojtak et al. 2008).

Due to projection effects, a fraction of the phase space is not accessible to observation. An observer
is able to measure velocity along the line of sight $v_{\rm los}$ and the position on the sky which can
be easily translated into the projected clustercentric distance $R$. This data set, when plotted
$v_{\rm los}$ versus $R$, is commonly referred to as the phase space diagram or the velocity diagram. Data
points in such diagrams are distributed according to the projected phase space density which is given by
(e.g. Dejonghe \& Merritt 1992; Merritt \& Saha 1993; Mahdavi \& Geller 2004)
\begin{equation}\label{f_los_def}
	f_{\rm los}(R,v_{\rm los})=2\pi R\int_{-z_{\rm max}}^{z_{\rm max}}\!\!\!\!\textrm{d}z\int\!\!\!\int_{E>0}\!\!\!\!
	\textrm{d}v_{R}\textrm{d}v_{\phi}f_{E}(E)f_{L}(L),
\end{equation}
where $z$ is the distance along the line sight from the cluster centre, while $v_{R}$ and $v_{\phi}$ are
velocity components in cylindrical coordinates. We will hereafter refer to $f_{\rm los}(R,v_{\rm los})$ as
the phase space density. The integration area of the last two integrals is given by the
circle $v_{R}^{2}+v_{\phi}^{2}<2\Psi(\sqrt{R^{2}+z^{2}})-v_{\rm los}^{2}$, where $\Psi(r)$ is a positively
defined gravitational potential for the NFW density profile, $\Psi(r)=V_{s}^{2}\ln(1+r/r_{s})/(r/r_{s})$
(Cole \& Lacey 1996). The boundaries of the integral along the
line of sight are given by the distance $\pm z_{\rm max}$ at which $v_{\rm
los}(R)$ becomes the escape velocity. It is worth mentioning that in (\ref{f_los_def}) we assume that the
density profiles of a tracer (galaxies) and dark matter are proportional to each other. Second, 
$f_{\rm los}$ is defined up to the normalization which must therefore be imposed by the following additional
condition
\begin{equation}	\label{df_los_norm}
	2\int_{0}^{R_{\rm max}}\!\!\!\textrm{d}R\int_{0}^{\sqrt{2\Psi(R)}}f_{\rm los}(R,v_{\rm
	los})\textrm{d}v_{\rm los}=1,
\end{equation}
where $R_{\rm max}$ is a cut-off radius of the phase space diagram.
From (\ref{df_los_norm}) one can immediately infer that the normalization factor is $\{r_{s}V_{s}[M_{\rm
los}(R_{\rm max})/M_{s}]\}^{-1}$, where $M_{\rm los}(R_{\rm max})$ is the projected mass within the aperture $R_{\rm max}$
for the NFW density profile (see {\L}okas \& Mamon 2001 for an analytical expression). From now on when
referring to the phase space density $f_{\rm los}$ we will always mean that it is properly normalized as
in (\ref{df_los_norm}).

We calculate the phase space density (\ref{f_los_def}) numerically using the algorithm of Gaussian
quadrature to evaluate each integral (Press et al. 1996). The energy part of the distribution function
$f_{E}(E)$ is interpolated between points which are equally spaced in energy and provide the numerical
solution of equation (\ref{rho_DF}). In order to remove improper boundaries of the integral along the line
of sight for $v_{\rm los}=0$ we changed the variable $z$ into $\Psi$. Next, to avoid singularity along the
line $L^{2}=(R^{2}+z^{2})v_{\phi}^{2}+(v_{R}z-v_{\rm los}R)^{2}=0$, when $\beta_{0}>0$, we used an even
number of abscissas for the variable $v_{\phi}$  so that $f_{L}(L)$ is never evaluated at $L=0$.

Fig.~\ref{df_los_ex} shows the contour maps of phase space density $f_{\rm los}(R,v_{\rm los})$ for 5
different anisotropy profiles plotted in the top right panel. In all cases we used $R_{\rm max}=5r_{s}$, a
typical virial radius of massive galaxy clusters, and the scale of transition between $\beta_{0}$ and
$\beta_{\infty}$, $L_{0}=0.2V_{s}r_{s}$, which corresponds to $\sim1r_{s}$. To facilitate
comparison, in the four bottom panels we plot the differences between the given and the isotropic
($\beta_{0}=\beta_{\infty}=0$) $f_{\rm los}$. The contours below the $0.01$ threshold are neglected as
insignificant for the typical number of data points considered in this work.

The results shown in Fig.~\ref{df_los_ex} are quite intuitive to interpret. We notice that $\beta_{0}$
modifies $f_{\rm los}$ in the central part of the diagram in such a way that radially biased models predict
more high velocity particles than tangentially biased ones (see the middle panels). On the other hand,
$\beta_{\infty}$ influences the outer part of a diagram so that the tangentially biased anisotropy
suppresses $f_{\rm los}$ at $v_{\rm los}=0$ and increases it for moderate velocities (see the bottom panels).
In the limit of $\beta_{\infty}\ll0$ the velocity distribution at $R/r_{s}\gg1$ takes the characteristic
horn-like shape with a minimum at $v_{\rm los}=0$ and a maximum at the circular velocity $\pm
[GM(R)/R]^{1/2}$ (see van der Marel \& Franx 1993 for a qualitative comparison). As a final remark on
Fig.~\ref{df_los_ex}, let us note that the parameters of the mass profile, $M_{s}$ and $r_{s}$, manifest
themselves only in the scaling of the axes of the diagram while preserving the shape of the phase space
density. This property, when put together with the way how $f_{\rm los}$ depends on the anisotropy,
automatically excludes the mass-anisotropy degeneracy that is an intrinsic problem of the
analysis based on the Jeans equation (see e.g. Merrifield \& Kent 1990).

\section{The mock catalogue of phase space diagrams}

The model of the phase space density introduced in the previous section provides an idealized description
of the real data. It neglects various secondary effects that occur in real galaxy clusters and perturb
their phase space diagrams. Among those the most important seem to be: the breaking of spherical symmetry,
the presence of substructures, the finite size of equilibrated zone and filamentary structures surrounding
the clusters and infalling towards them.
%The last two effects, which are a consequence of
%a simple fact that clusters are currently being formed, are probably the most significant, because
%it turns out that the real phase space diagrams are unavoidably populated by some fraction of galaxies
%which are not in equilibrium with the other parts of the clusters.
In order to study the impact of such effects on the reliability of our approach to data analysis we need
to construct a mock catalogue of phase space diagrams which on the one hand resembles the real spectroscopic
cluster survey and on the other provides the true mass and anisotropy profiles. This can be easily achieved
with the use of cosmological simulations.

\begin{figure}
\begin{center}
    \leavevmode
    \epsfxsize=8cm
    \epsfbox[50 50 580 230]{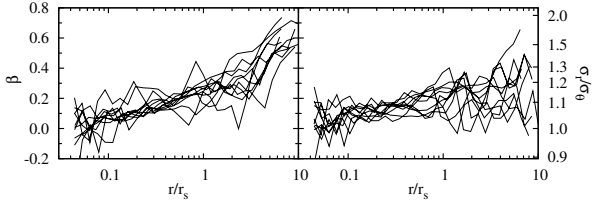}
\end{center}
\caption{The anisotropy profiles in 20 dark matter haloes selected from the simulation. The two
panels show separately the profiles for a subsample of ten haloes with more steeply (left) and less steeply
(right) increasing profiles.}
\label{beta_3d}
\end{figure}

We used an $N$-body cosmological simulation of a $\Lambda$CDM model. The simulation was carried out in a box
of $160\hMpc$ with WMAP3 cosmological parameters (Spergel et al. 2007): $\Omega_{m}=0.24$, $\Omega_{\Lambda}=0.76$ and
the dimensionless Hubble constant $h=0.73$ (for more details see Wojtak et al. 2008). We identified all cluster-size
haloes at redshift $z=0$ and selected $20$ of them that appeared not to be products of recent major mergers.
For each halo we measured the virial mass $M_{v}$ and the virial radius $r_{v}$ which are defined in
terms of the mean density within the virial sphere by the following equation
\begin{equation}
	\frac{M_{v}}{(4/3)\pi r_{v}^{3}}=\Delta_{c}\rho_{c},
\end{equation}
where $\Delta_{c}$ is the so-called overdensity parameter. For cosmological model under consideration we
adopted $\Delta_{c}=94$ (see e.g. {\L}okas \& Hoffman 2001). The virial mass of selected haloes
is in the range $(1.5-20)\times 10^{14}\Msun$. The minimum number of particles
inside the virial sphere is $4.2\times 10^{5}$.

By fitting the NFW formula (\ref{NFW}) to the density profile calculated in radial bins equally spaced in
logarithmic scale we determined the scale radius $r_{s}$ of each halo and then the concentration parameter
$c=r_{v}/r_{s}$. The scale mass $M_{s}$ was measured directly as the mass enclosed by the sphere of radius
$r_{s}$. We found that the mass profile extrapolated from $r_{s}$ outward to the virial sphere recovers the
virial mass with the precision of a few percent.

Fig.~\ref{beta_3d} shows the anisotropy profiles in selected dark matter haloes. The profiles were
measured from velocity dispersions of dark matter particles inside thin spherical shells of radii extending
to $1.5r_{v}$. Although a general trend for radial variation of the anisotropy is very clear, the profiles are
considerably scattered. One can distinguish at least two classes of anisotropy profiles: steeply rising and
flat or moderately rising. According to this criterion we divided our halo sample into two equally numerous
groups of ten haloes each (see the left and right panel of Fig.~\ref{beta_3d}). The reason for this division
will become clear in section 5, where we consider constraints on the anisotropy profile from the joint
analysis of many phase space diagrams.

In order to construct phase space diagrams associated with the haloes we pick a line of sight and
find all particles inside a cylinder of observation. The selection is restricted to the particles 
with the projected halocentric distances $R<r_{v}$ and velocities along the line of sight
$v_{\rm los}$ (with the Hubble flow included) in the range $\pm 4000$ km s$^{-1}$ in the rest frame of 
a halo, which is the commonly used and sufficiently conservative velocity cut-off. 
%Velocities $v_{\rm los}$ include the Hubble flow as in real
%observations. 
The final phase space diagrams are generated by drawing randomly $300$ particles for each halo,
where we assume, to make the scheme unambiguous, that the tracer is given simply by the particles. The number
of data points is fixed at a typical number of spectroscopic redshifts available for a nearby galaxy cluster
with the same criterion of velocity cut-off. The selection of particles is done three times for three lines
of sight chosen to be parallel to the Cartesian axes of the simulation box. Thus we effectively get three
independent sets of phase space diagrams for the same sample of 20 haloes.

Due to projection effects, the phase space diagrams include the so-called interlopers: the particles
(galaxies) of background or foreground which are not physically connected to the haloes (clusters). It is
commonly accepted that any dynamical analysis should be necessarily preceded by an identification and
removal of as many of these objects as possible.
In this work we apply the procedure proposed by den Hartog \& Katgert (1996) which appears to be one of the
most effective algorithms for interloper removal (Wojtak et al. 2007; Wojtak \& {\L}okas 2007). The method
is based on an iterative rejection of galaxies with velocity exceeding a maximum velocity which is a function
of the clustercentric distance $R$. The maximum velocity is calculated using a model which
allows the galaxies to be either on circular orbit with velocity $v_{\rm cir}=\sqrt{GM(r)/r}$ or to fall
towards the cluster centre with velocity $\sqrt{2}v_{\rm cir}$, where $M(r)$ is approximated in each
iteration by the virial mass estimator (Heisler, Tremaine \& Bahcall 1985). The phase space diagrams
preprocessed in such a way are suitable for the proper analysis in terms of the phase space density.
Hereafter, when referring to the phase space diagram we will always mean the diagram after interloper removal.

\section{Bayesian analysis and MCMC}

Our aim is to constrain the parameters of the mass and anisotropy profiles by analyzing
the mock phase space diagrams generated from the simulation. Following the principle of Bayesian inference our
knowledge of model parameters ${\mathbi a}$ given the data $D$ may be expressed as the posterior probability
$p({\mathbi a}|D)$ which is related to the probability of obtaining the data in a given model
$p(D|{\mathbi a})$ (likelihood) and the prior probability $p({\mathbi a})$ via equation
\begin{equation}
	p({\mathbi a}|D)=\frac{p({\mathbi a}) p(D|{\mathbi a})}{p(D)},
\end{equation}
where ${\mathbi a}$ is a vector of model parameters and $D$ stands for the data. The probability
$p(D)$ plays the role of the normalization coefficient and can be neglected without loss of
generality.

The data $D$ of a phase space diagram consists of $N$ points located at $(R_{i},v_{{\rm los},i})$, where
$i=1,\dots,N$. Assuming that the number and mass density are proportional to each other the likelihood 
is given by
\begin{equation}	\label{like_1diagram}
	p(D|{\mathbi a})= \prod_{i=1}^{N} f_{\rm los}(R_{i},v_{{\rm los},i}|{\mathbi a}),
\end{equation}
where $f_{\rm los}$ is the phase space density given by (\ref{f_los_def}) and 
${\mathbi a}=\{M_{s},r_{s},\beta_{0},\beta_{\infty},L_{0}\}$. 
It is worth mentioning that this formula does not take into account statistical errors of $v_{\rm los}$ 
and $R$. This is motivated by the fact that the errors of spectroscopic redshift measurements for nearby 
clusters are quite small and have negligible impact on the results of the analysis (see e.g. van der Marel 
et al. 2000).

\subsection{Priors and reparametrization}

The two parameters of the mass profile, $M_{s}$ and $r_{s}$, are the scale parameters. The most appropriate
prior probability for such parameters is the Jeffreys prior which represents equal probability
per decade (Jeffreys 1946), i.e. $p(M_{s})\propto 1/M_{s}$ and $p(r_{s})\propto 1/r_{s}$. We will use
this kind of prior in the analysis of any single phase space diagram. The situation will change
in subsection $5.3$ where we consider joint analysis of many phase space diagrams.

The commonly used anisotropy parameter $\beta(r)$ gives rise to unequal weights for tangentially and
radially biased models. In order to make the prior probability for both types of anisotropy equal we
introduce, following Wilkinson et al. (2002) and Mahdavi \& Geller (2004), the following
reparametrization
\begin{equation}	\label{beta_repar}
	\ln \frac{\sigma_{r}(r)}{\sigma_{\theta}(r)}=-\frac{1}{2}\ln[1-\beta(r)].
\end{equation}
Using a uniform prior for $\ln(\sigma_{r}/\sigma_{\theta})_{0}=-1/2\ln(1-\beta_{0})$ and
$\ln(\sigma_{r}/\sigma_{\theta})_{\infty}=-1/2\ln(1-\beta_{\infty})$ we put equal weights
to all types of anisotropy. The ranges of both priors are limited by $\beta_{0}\le 1/2$ and
$\beta_{\infty}\le 0.99$. The first condition is motivated by the requirement of the
distribution function to be positive (see An \& Evans 2006), whereas the second one allows us to avoid
improper posterior distribution that is important when the data favour $\beta_{\infty}\lesssim 1$.

The scale of transition between the two asymptotic values of anisotropy is defined by $L_{0}$. We
find that keeping this parameter free causes degeneracy with $\beta_{\infty}$ and $\beta_{0}$.
This considerably restricts the information on the growth or decrease of the anisotropy profile
that $\beta_{0}$ and $\beta_{\infty}$ are expected to convey. To overcome this difficulty we fix
$L_{0}$ at $0.2V_{s}r_{s}$, a value corresponding to the $\sim 1r_{s}$ transition scale
expected from cosmological simulations (see Wojtak et al. 2008). This implies that the anisotropy
profiles approach their asymptotic values at some characteristic scales of radius,
namely $\beta_{0}=\beta(r\lesssim 0.1r_{s})$ and $\beta_{\infty}=\beta(r\gtrsim 10r_{s})$. 
Fixing $L_{0}$ reduces the dimension of the parameter space by one so that the analysis is based 
on the four-parameter phase space model specified by $M_{s}$, $r_{s}$, $\beta_{0}$ and 
$\beta_{\infty}$.

\subsection{MCMC approach}

Our purpose is to calculate the posterior probability density and provide credibility regions in the
parameter space. This involves numerical integrations of multivariate semi-Gaussian functions that
can be efficiently tackled with the Markov Chain Monte Carlo (MCMC) algorithm. The main idea of this
approach is to construct a sufficiently long chain of models which are distributed in the parameter
space according to the posterior probability density. Once such a chain is provided, one can easily
compute the marginal probability distributions by projecting all points on to an appropriate parameter
subspace and evaluating the histograms.

We construct the Markov chains following the Metropolis-Hastings algorithm (see e.g. Gregory 2005).
In the first step of this algorithm one picks a trial point in parameter space ${\mathbi a}_{t+1}$
by drawing from the so-called proposal distribution $q({\mathbi a}|{\mathbi a}_{t})$ centred on the
previous point of the chain ${\mathbi a}_{t}$. Then the point ${\mathbi a}_{t+1}$ is accepted with the
probability equal to $\textrm{min}\{1,r\}$, where $r$ is the so-called Metropolis ratio given by
\begin{equation}
	r=\frac{p({\mathbi a}_{t+1}|D)q({\mathbi a}_{t}|{\mathbi a}_{t+1})}
	{p({\mathbi a}_{t}|D)q({\mathbi a}_{t+1}|{\mathbi a}_{t})},
\end{equation}
otherwise it takes the value of its predecessor, ${\mathbi a}_{t+1}={\mathbi a}_{t}$. In the case of a few
parameter model the proposal probability density can be any function that roughly resembles the target
posterior distribution. In our work we use a multivariate Gaussian with a diagonal covariance matrix.
The variances are calculated from short trial Markov chains of about $2000$ models, where our best
initial guess for parameter variances is used. This approach is a simplified version of the scheme
outlined in Widrow, Pym \& Dubinski (2008). It is worth mentioning that the proposal distribution
is symmetric, $q({\mathbi a}_{t+1}|{\mathbi a}_{t})=q({\mathbi a}_{t}|{\mathbi a}_{t+1})$,
so that the Metropolis ratio is given by $p({\mathbi a}_{t+1}|D)/p({\mathbi a}_{t}|D)$. 
%We note that our final Markov chains achieve the so-called acceptance rate (the average rate at
%which proposed models are accepted) of $20-30$ percent that is a recommended calibration for many-parameter
%models (see e.g. Gregory 2005).

An important issue of the MCMC analysis is a relative number of distinct points in the parameter space 
which is described by the so-called acceptance rate (the average rate at which proposed models are accepted). 
A recommended value of this parameter for many-parameter models is around $20-30$ percent (see e.g. Gregory 2005). 
We note that our choice of the proposal distribution keeps the acceptance rates of the resulting Markov chains 
within this range. Some additional modification of the proposal distribution is required in subsection 
5.3, where we consider a joint analysis involving $N_{p}=22$ parameters. In order to keep the acceptance 
rate at a desirable level all initial variances of the proposal distribution are scaled by $2.4^{2}/N_{p}$ 
(see e.g. Gelman et al. 1995). This correction prevents the acceptance rate from dropping, due to a large 
number of parameters in use, to a very low value of around 1 percent.

\begin{figure}
\begin{center}
    \leavevmode
    \epsfxsize=8cm
    \epsfbox[50 50 590 800]{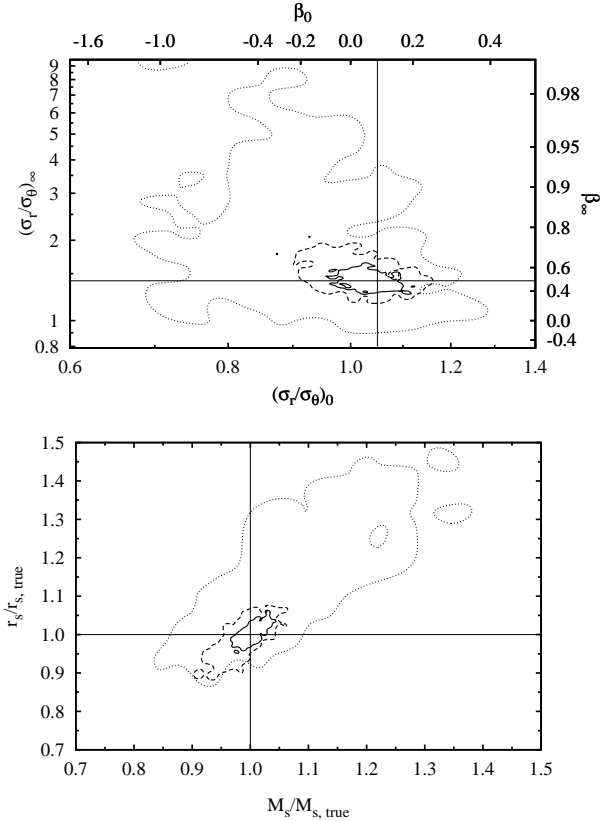}
\end{center}
\caption{The $1\sigma$ credibility regions inferred from the MCMC analysis of three
theoretical phase space diagrams generated from the distribution function with
$\beta_{0}=0.1$ and $\beta_{\infty}=0.5$. The parameters of the mass profile
were rescaled by their true values. Solid, dashed and dotted lines correspond
to the diagrams with $9000$, $3000$ and $300$ data points. Straight lines indicate
the true values of the parameters.}
\label{1sig_3dia}
\end{figure}

When applying the MCMC algorithm, it is crucial to check whether the chains explore properly
the parameter space, i.e. whether they possess the property often referred to as mixing. To
save computing time, we decided not to create several chains for each data sample that is a commonly
advised way to look for convergence. As an alternative, we follow two simple indicators of mixing.
First, we calculate parameter dispersions in two halves of a given chain and within the whole chain. If the
relative differences between them do not exceed 10 percent for each parameter, the chain is considered 
to be mixed. Second, we monitor the variation of the posterior probability
along the chains. Unless the profiles of $p({\mathbi a}|D)$ exhibit a long-scale tendency to grow or decline,
the chain is again expected to be mixed. We note that the so-called burn-in part of the chains, when the
first models gradually approach the most favoured zone of the parameter space, is not longer than 1 percent
of the total length of the chains. All chains used in this analysis consist of $10^{4}$ models
that is above the recommended minimum of $330 N_{p}$ (Dunkley et al. 2005).

\subsection{Number of data points}

One of the leading factors affecting the posterior probability is the limited number of
data points. In order to study the impact of this effect we carried out the analysis
of three theoretical phase space diagrams with $300$, $3000$ and $9000$ points
that correspond to a typical size of a data sample for a single nearby galaxy cluster
(the first number) and to a compilation of $10-30$ of them (the last two numbers).
The diagrams were generated from a discrete representation of the full phase space
distribution function with the following parameters of the anisotropy profile:
$\beta_{0}=0.1$, $\beta_{\infty}=0.5$ and $L_{0}=0.2V_{s}r_{s}$. The phase space
was sampled using the acceptance-rejection technique (Press et al. 1996)
in a manner described by Kazantzidis, Magorrian \& Moore (2004).

\begin{figure}
\begin{center}
    \leavevmode
    \epsfxsize=8cm
    \epsfbox[50 50 590 420]{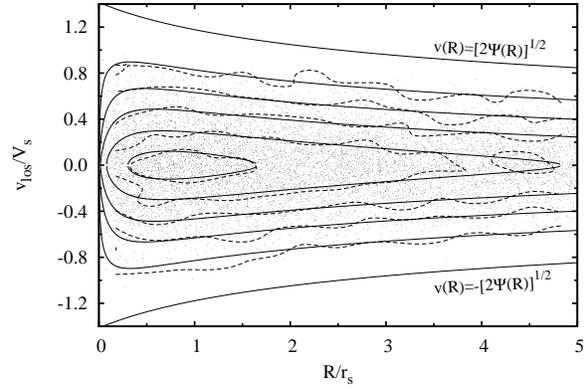}
\end{center}
\caption{Projected phase space diagram obtained from the full phase space distribution
function with anisotropy parameters $\beta_{0}=0.1$ and $\beta_{\infty}=0.5$.
Solid and dashed lines show respectively the isodensity contours of the best-fitting model
and the smoothed surface density of $9000$ sampling points. Two envelope lines
labelled by $v(R)=\pm\sqrt{2\Psi(R)}$ are the profiles of the escape velocity.}
\label{9k_cont}
\end{figure}

Fig.~\ref{1sig_3dia} shows the $1\sigma$ credibility regions, i.e. the regions enclosing $68.3$ percent of the 
corresponding marginal probability, inferred from the MCMC analysis of the three theoretical diagrams. 
For the sake of simplicity, the parameters of
the mass profile were rescaled by the true values. The best-fitting model for $N=9000$ case
is overplotted in solid lines on top of the smoothed contour map of the phase space density (dashed lines)
in Fig.~\ref{9k_cont}. From Fig.~\ref{1sig_3dia} we conclude that typical relative errors of $M_{s}$ and
$r_{s}$ from the analysis of a single phase space diagram with $N=300$ points are about 20 percent. On the other hand, the
corresponding constraints on the anisotropy parameters are very poor. The marginal distribution is so wide,
particularly along $\beta_{\infty}$, that it is expected to be sensitive to any kind of noise in the
real data. Anticipating the results of the following section we emphasize that the only way to reliably
constrain the anisotropy profile is to increase the number of data points. In practice, this can be achieved
in the joint analysis of at least $10$ independent diagrams ($3000$ data points), where one assumes a
universal anisotropy profile. The result for $N=3000,9000$ in the top panel of Fig.~\ref{1sig_3dia} shows
what we can expect from such analysis.

\section{Results of the analysis}

We carried out the MCMC analysis of $60$ phase space diagrams of our mock data
catalogue described in section 3. The results are presented in the form of the
maximum a posteriori (MAP) values of the parameters and the errors that correspond
to the boundary of the $1\sigma$ credibility regions of the marginal distributions, 
i.e. the regions enclosing $68.3$ percent of the marginal probability.

\begin{figure}
\begin{center}
    \leavevmode
    \epsfxsize=8cm
    \epsfbox[55 55 580 1050]{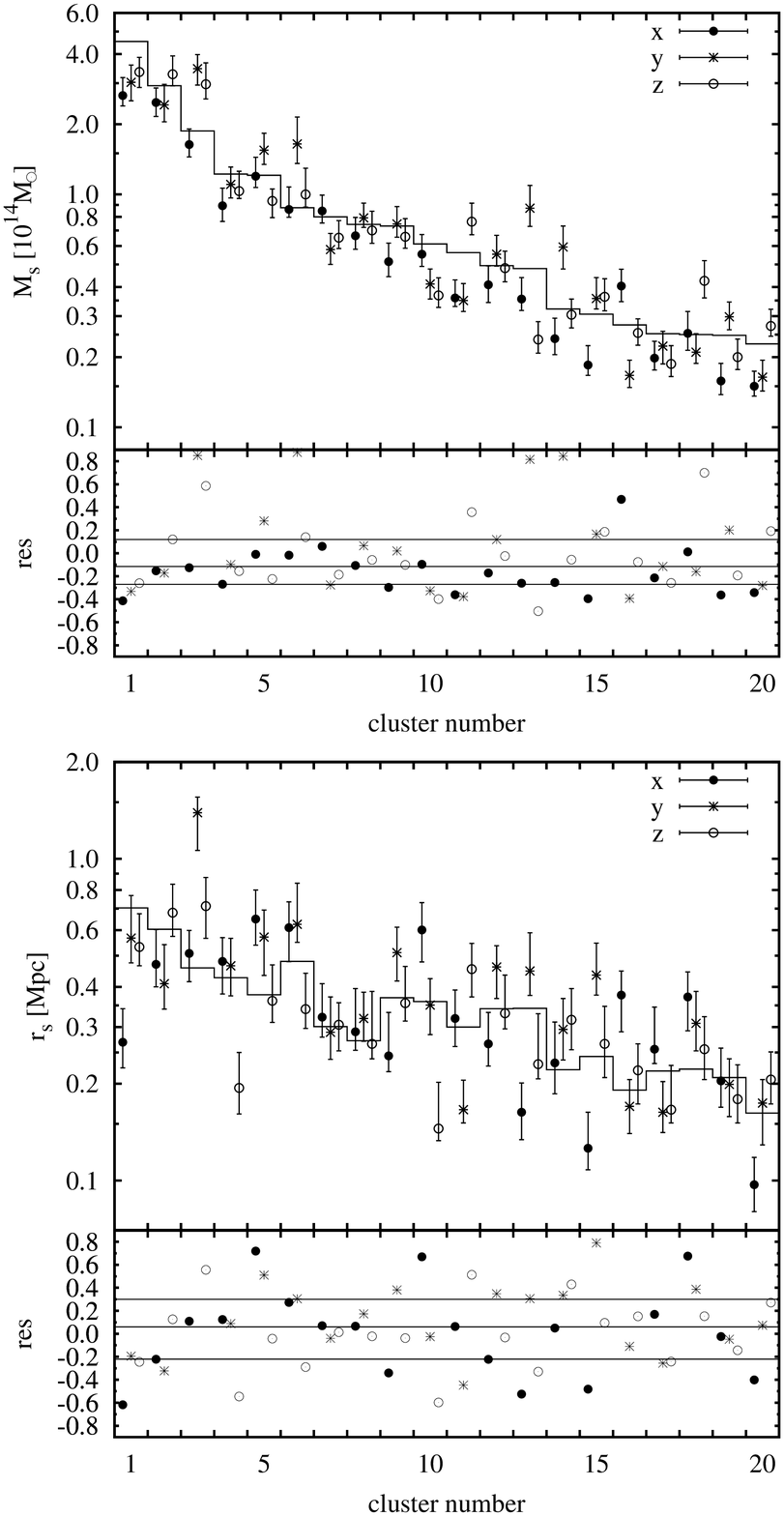}
\end{center}
\caption{Mass profile parameters obtained from the MCMC fits of 20 individual simulated clusters
observed in 3 directions. The symbols indicate MAP (maximum a posteriori) values and the errors correspond
to the boundary of the $1\sigma$ credibility region of the marginal distributions. Filled circles, asterisks
and empty circles refer to directions of projection along the $x$, $y$ and $z$ axis of the simulation box.
Solid broken lines indicate the true values of the parameters. The three lines in the lower panels showing
fractional residuals from the true parameter values indicate the quartiles of a total set of $60$ residuals.
The clusters are ordered by decreasing scale mass $M_{s}$.}
\label{msrs_60}
\end{figure}

\begin{figure}
\begin{center}
    \leavevmode
    \epsfxsize=8cm
    \epsfbox[55 55 580 1050]{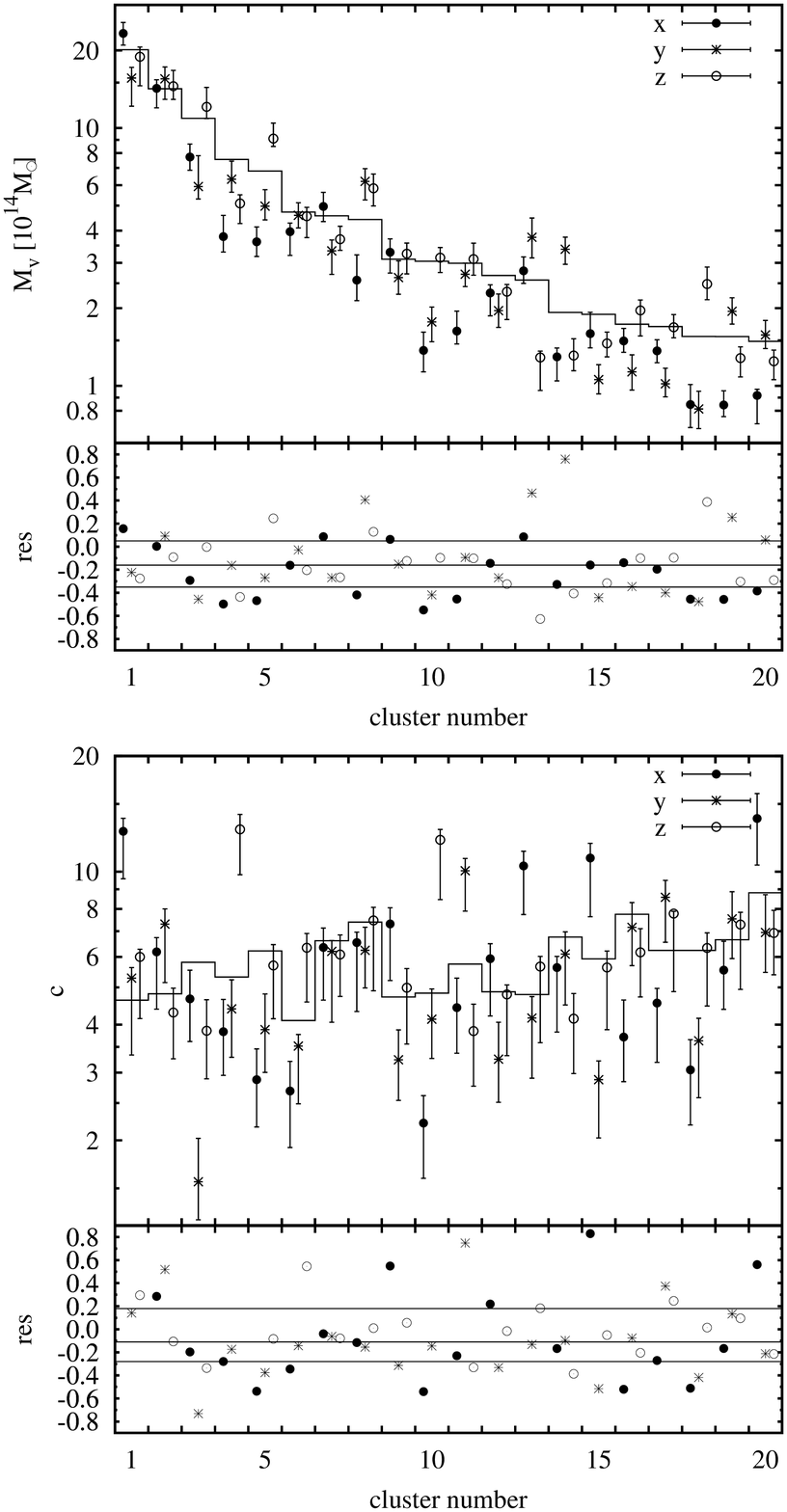}
\end{center}
\caption{Same as Fig.~\ref{msrs_60}, with the scale mass and radius converted to the virial mass $M_{v}$
and concentration $c$. The clusters are ordered by decreasing virial mass.}
\label{mvc_60}
\end{figure}

\subsection{Mass profile}

Fig.~\ref{msrs_60} shows the results for the parameters of the mass profile, $M_{s}$
and $r_{s}$. From the diagrams showing residuals, attached below the mains plots, we conclude
that in half of the cases the relative errors do not exceed 25 percent (see the lines of the first
and third quartile of the residuals). On the other hand, around 15 percent of the results 
differ from the real values by more than 50 percent that can hardly be reconciled 
with the expectations from the effect of a shot noise. Analyzing a few tens of theoretical 
phase space diagrams from subsection 4.3 with $N=300$ data points we find that residuals 
induced by a shot noise do not exceed 30 percent. This may suggest that the most outlying points in 
Fig.~\ref{msrs_60} are probably subject to non-negligible systematic errors. 
Potential sources of these errors could include projection effects
as well as those associated with internal structure. We looked for the influence of
the shape of dark matter haloes and found no correlation between the accuracy of parameter
estimates and the halo ellipticity or the alignment of the major axis with respect to the line
of sight.

%In order to estimate the magnitude of systematic errors we calculated dispersions of the MAP
%values with respect to the true ones. We found that in both cases, for $M_{s}$ and $r_{s}$,
%this dispersion is about twice bigger than the typical dispersion of the marginal
%distribution.
%Therefore the expected systematic error exceeds the statistical one by
%factor $1.7$.
%We emphasize, however, that only the most outlying results are affected by systematic errors.

\begin{figure}
\begin{center}
    \leavevmode
    \epsfxsize=7.8cm
    \epsfbox[60 60 580 1180]{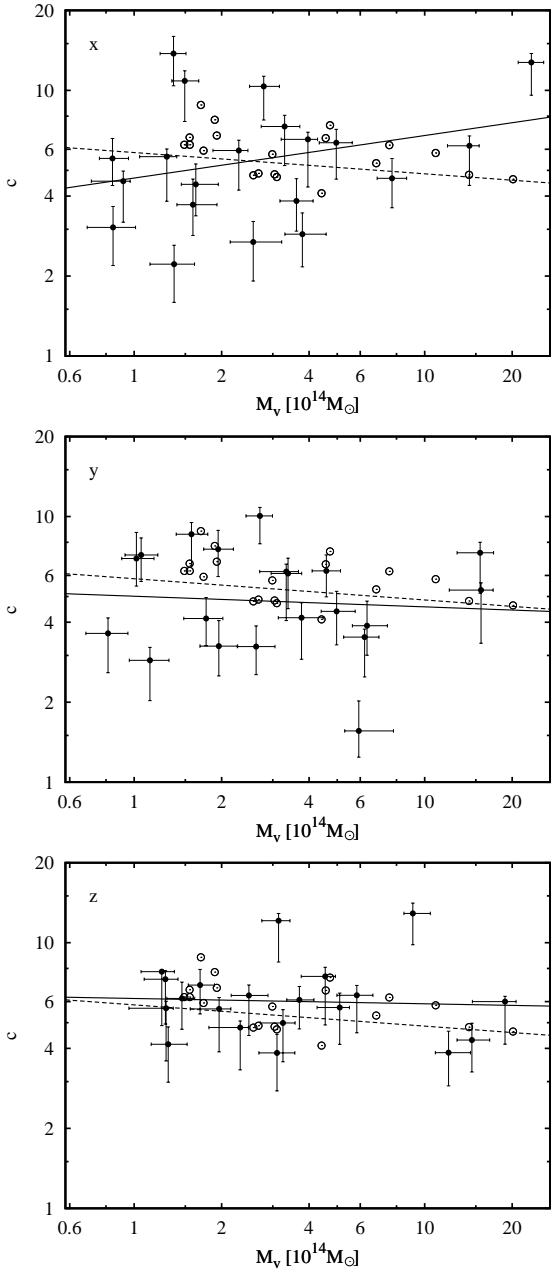}
\end{center}
\caption{Prospects of reconstructing the mass-concentration relation, based on our MCMC analysis
of 20 individual simulated clusters observed in 3 directions labelled by $x$, $y$ and $z$ (from the top to the bottom
panel).
The solid lines are the power-law fits to the best-fitting parameters obtained from the analysis of
the phase space diagrams (filled circles). Open circles indicate the true values of the parameters for our
20 clusters. The dashed line repeated in each panel is a power-law fit to the mean mass-concentration
relation for WMAP3 cosmology from Macci\`o et al. (2008).}
\label{c-m_pers}
\end{figure}

All Markov chains can be easily transformed into the chains of models parametrized by
the virial mass $M_{v}$ and the concentration $c$. The resulting credibility ranges and the
MAP values of these parameters are shown in Fig.~\ref{mvc_60}. We find that the scatter in
residuals is comparable to the case where the scale parameters were used and the positions of the most
discrepant points imply the presence of similar systematic errors. There is noticeable progress
in the constraints on the concentration parameter in comparison to previous work based on
modelling velocity moments (Sanchis et al. 2004; {\L}okas et al. 2006). On the other hand, there
is no convincing evidence for a similar improvement for the virial mass.

The MAP values of the mass parameters typically underestimate the true values by 15 percent and 11 percent
for the virial and scale mass respectively (see the position of median lines in the diagrams showing
residuals in Figs.~\ref{mvc_60} and \ref{msrs_60}). Interestingly, a similar offset was reported by
Biviano et al. (2006) for the virial mass estimated from the Jeans analysis of velocity dispersion profiles.
They suggested that this bias is related to the presence of interlopers infalling
towards the cluster along filaments. Due to relatively small velocities in the rest frame of a cluster,
these objects cannot be identified by any algorithm of interloper removal and, therefore, remain in the
sample decreasing the velocity dispersion and the resulting virial mass. We think that a
similar mechanism is likely responsible for the offset of our results. Nevertheless, we emphasize
that this bias is smaller than the statistical errors obtained in the MCMC analysis so that the overall effect
is not statistically significant.

It is a well known fact the concentration parameter is weakly correlated to the virial mass
(e.g. Navarro et al. 1997;
Bullock et al. 2001). This so-called mass-concentration relation is an imprint of the formation history and is therefore
thought to be one of the predictions of the cosmological model. Fig.~\ref{c-m_pers} demonstrates the
prospects of recovering this relation with our approach. Each panel in this Figure
shows the results obtained for 20 phase space diagrams for a given direction of observation (filled circles
with error bars) and the best power-law fit (solid line). For comparison with the prediction of the
$\Lambda$CDM model, we plot the true parameters of our 20 clusters (open circles) and with a dashed line the
power-law fit to the mean mass-concentration relation for relaxed haloes simulated in the framework of WMAP3
cosmology from Macci\`o, Dutton \& van den Bosch (2008).
We find that, within errors, the empirical $M_{v}$-$c$ relation is consistent
with a flat profile in all cases. For one projection (top panel) the slope deviates from the expected value of $-0.1$
by more than two sigma. The normalization within the mass range under consideration does not exhibit any offset with
respect to the prediction so that future determination of this quantity for real clusters seems to be
feasible. This result is particularly important in the context of recently discussed inconsistency between
the normalization from the simulations and observational constraints (see e.g. Comerford \& Natarajan 2007).

\begin{figure}
\begin{center}
    \leavevmode
    \epsfxsize=8cm
    \epsfbox[55 55 580 700]{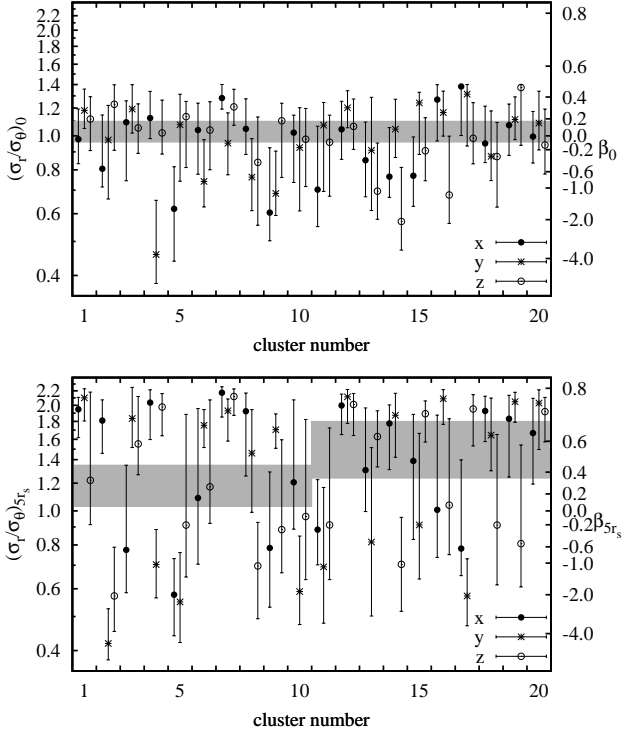}
\end{center}
\caption{Anisotropy profile parameters obtained from the MCMC fits of 20 individual simulated clusters observed
in 3 directions.
%Results of the MCMC analysis for the anisotropy parameters: the asymptotic
%anisotropy in the cluster centre, $\beta_0$ and the anisotropy at clustercentric distance $5r_{s}$, $\beta_{5r_{s}}$.
%Both parameters were also expressed in terms of velocity dispersions, $(\sigma_{r}/\sigma_{\theta})_{0}$ and
%$(\sigma_{r}/\sigma_{\theta})_{5r_{s}}$.
Gray boxes indicate the range of the true anisotropy at $0.1r_{s}$
(top panel) and $5r_{s}$ (bottom panel). The first and second ten clusters correspond to the samples with
flat to moderately rising and steeply rising anisotropy profiles respectively, as shown in
Fig.~\ref{beta_3d}. The meaning of all symbols is the same as in Fig.~\ref{msrs_60}. }
\label{ani_60}
\end{figure}

\subsection{The anisotropy profile}

Fig.~\ref{ani_60} shows the results of the MCMC analysis for the anisotropy parameters. Since
the phase space model is well-defined only within the virial sphere, we decided to replace the parameter
$\beta_{\infty}$ with $\beta_{5r_{s}}$ measured at $5r_{s}$ which is the typical
scale of the virial radius. This allows us to avoid extrapolation of the model
beyond the virial sphere. For comparison the Figure also shows in gray boxes the range of anisotropy values
at $0.1r_{s}$ and $5r_{s}$ in two samples of clusters introduced is section $3$ which differ in steepness of
their anisotropy profiles.

The majority of results for $\beta_{0}$ are consistent with the true values. The errors are quite large,
however, so they would not allow to distinguish between radially and tangentially biased anisotropy. We
emphasize that all $1\sigma$ credibility ranges are clearly detached from the upper boundary of the prior
probability, $\beta=0.5$. This means that this limit is not artificial but is embedded in the phase space
structure of an equilibrated system.

\begin{figure}
\begin{center}
    \leavevmode
    \epsfxsize=8cm
    \epsfbox[55 55 580 810]{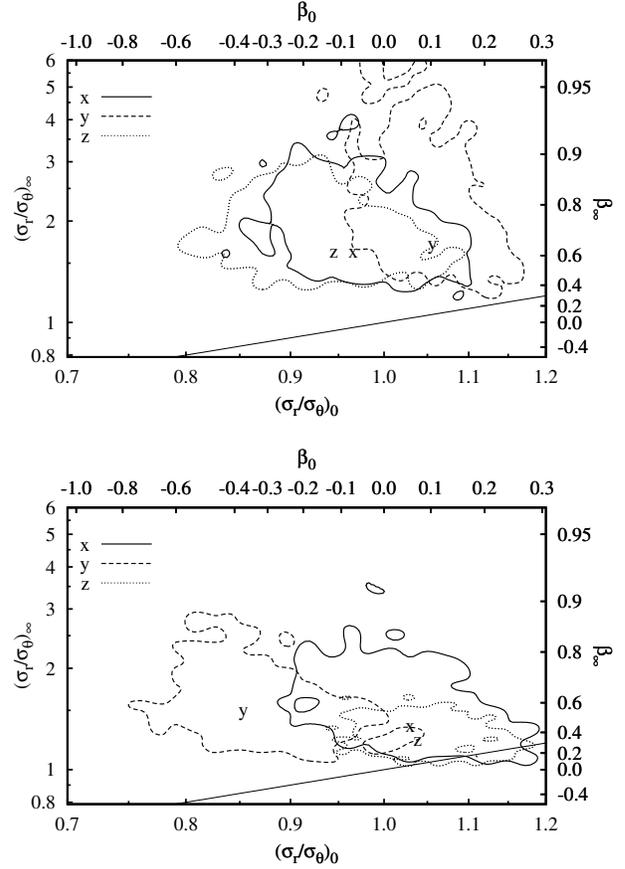}
\end{center}
\caption{Anisotropy profile parameters obtained from the joint MCMC fits of 10 simulated clusters
with rapidly (top panel) and moderately (bottom panel) rising anisotropy profiles. The contours indicate
the boundaries of the $1\sigma$ credibility regions with solid, dashed and dotted lines referring to the projection
along the $x$, $y$ and $z$ axis respectively. The corresponding maximum a posteriori (MAP) values are indicated with
letters $x$, $y$ and $z$. The straight solid line represents a family of flat anisotropy profiles.
}
\label{ani_joint}
\end{figure}

The anisotropy at $5r_{s}$ is poorly constrained. Although there is a weak tendency for the results to
cluster around the true values, most of them are shifted considerably upwards or downwards. An important
circumstance responsible for this bias is the fact that the posterior probability distribution is so wide
in the space of anisotropy parameters that it is sensitive to any irregularities in the phase space
diagram. These irregularities often occur in the outermost part of the diagrams, mostly due to subclustering
and the presence of small velocity interlopers from the outside of the virial sphere. The overall consequence
is that the posterior probability is often perturbed in $\beta_{\infty}$ (or $\beta_{5r_{s}}$) and typically
stable in $\beta_{0}$.
Due to significant systematic errors of $\beta_{5r_{s}}$ (or $\beta_{\infty}$), the constraints on
the radial variation of the anisotropy for a single galaxy cluster are very weak. Most probably this
caveat cannot be avoided in any approach aiming to determine the anisotropy profile of
galaxies in a cluster (see e.g. Hwang \& Lee 2008).

\subsection{The anisotropy profile from the joint analysis}

The only possibility to obtain reliable constraints on the anisotropy profile is to increase the number
of data points used in the analysis. In subsection 4.3 we showed that a sample of $3000$ data points
is expected to provide conclusive results in this matter. When dealing with galaxy clusters, such
increase in the amount of data can only be achieved at present by means of the joint analysis of several
clusters ($\sim 10$ clusters). An additional advantage of this approach is that all local irregularities of
the phase space diagrams compensate each other minimizing systematic errors.

A common approach to carry out a joint analysis of several clusters consists of two stages.
First, one merges all phases space diagrams with properly rescaled radii $R$ and velocities
$v_{\rm los}$. As the units of phase space coordinates one most often uses the virial radius $r_{v}$
and the so-called virial velocity $V_{v}=(GM_{v}/r_{v})^{1/2}$. Then, assuming homology of the cluster
sample, the resulting phase space diagram is reanalyzed in the framework of a model reduced by the
parameters that were used to scale the radii and velocities. The final constraints obtained in this
approach are thought to represent typical properties of galaxy clusters in a given sample
(e.g. van der Marel et al. 2000; Mahdavi \& Geller 2004; {\L}okas et al. 2006).

Our approach differs from the above scheme in both respects. First, instead of merging many phase space
diagrams we introduce the following generalization of the likelihood function (\ref{like_1diagram}) for the
joint analysis of $n$ phase space diagrams
\begin{eqnarray}
	f(M_{s,1},\dots,M_{s,n},r_{s,1},\dots,r_{s,n},\beta_{0},\beta_{\infty})=\nonumber\\
	\prod_{j=1}^{n}\prod_{i=1}^{N}f_{\rm los}(R_{j,i},v_{{\rm los},j,i}|\{M_{s,j},r_{s,j},
	\beta_{0},\beta_{\infty}\})		\label{like_ndiagram},
\end{eqnarray}
where $j$ and $i$ are respectively the reference number of a cluster and a data point of $j$-th phase
space diagram. The likelihood as well as the posterior probability distribution are functions of $2n+2$
parameters (the parameter $L_{0}$ is fixed as discussed in subsection 4.1). We assume that the two parameters
of the anisotropy profile are common to all clusters so that the final result concerning both of them is
expected to provide a typical variation of the anisotropy in a given cluster sample. Our knowledge of
$M_{s,i}$ and $r_{s,i}$ obtained in the previous analysis (see Fig.~\ref{msrs_60}) is incorporated in the
prior probability. For the sake of maintaining the analytical form of the prior, we used the product of $2n$
Gaussian distributions centred at the MAP values of $M_{s,i}$ and $r_{s,i}$. The dispersion of each Gaussian
was fixed at the dispersion of the corresponding marginal probability distribution. It is interesting to note
that if we arbitrarily ignored the credibility region of the nuisance parameters $M_{s,i}$ and $r_{s,i}$,
our approach would be conceptually very similar to the case when all phase space diagrams are merged except
that radii and velocities would be scaled by $r_{s}$ and $V_{s}$.

\begin{figure}
\begin{center}
    \leavevmode
    \epsfxsize=8cm
    \epsfbox[55 55 580 610]{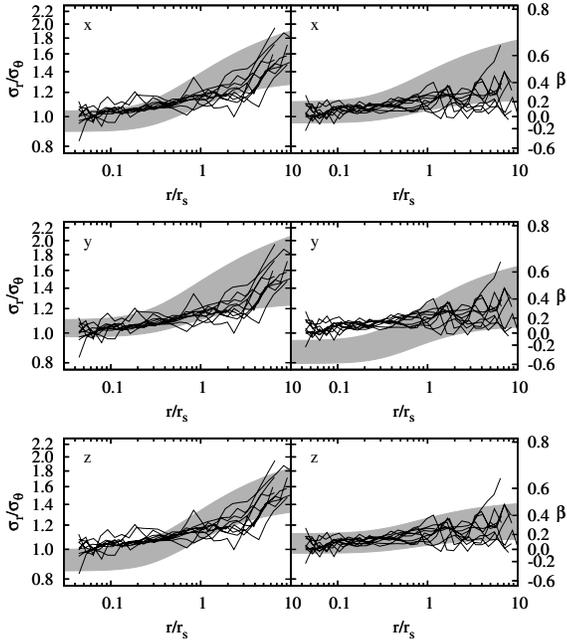}
\end{center}
\caption{Anisotropy profile obtained from the joint MCMC fits of 10 simulated clusters with
rapidly (left panels) and moderately (right panels) rising anisotropy profiles, in three
directions of observation (rows). The widths of the shaded regions are given by the $1\sigma$
credibility range of the anisotropy at a given radius. Solid lines are the true anisotropy profiles
measured from the simulations.
}
\label{ani_prof}
\end{figure}

We analyzed two samples of clusters, with $n=10$ objects each, observed in three directions as described in
section 3. The samples separate the objects with distinctly rising anisotropy profiles from those with
flat or moderately rising ones. This provides an opportunity to verify whether our approach is able to
distinguish between these two cases. Fig.~\ref{ani_joint} shows the $1\sigma$ credibility regions in the
$\beta_{0}-\beta_{\infty}$ plane inferred from the Markov chains of $10^{4}$ models. First, we notice that
all results are consistent with the velocity dispersion tensor that is isotropic in the cluster centre and
radially biased outside. Second, the resulting anisotropy profiles are clearly steeper for clusters with
rising anisotropy profiles (top panel) and flatter for the others (bottom panel).

In order to check the radial variation of the anisotropy in Fig.~\ref{ani_prof} we plot the $1\sigma$
credibility regions of its radial profiles. The resulting anisotropy effectively traces the true
profiles (solid lines) within the radial range covering more than two orders of magnitude around $r_{s}$.
A local deviation from the true values occurs only once (middle right panel for $r<0.5r_{s}$) as a result of
relatively low quality of this data sample (see the results shown with asterisks in Fig.~\ref{ani_60}).
Since the constraints on $\beta_{0}$ are considerably tighter than on $\beta_{\infty}$, statistical errors of
local anisotropy typically increase with radius. Nevertheless, it is still possible to distinguish between
steeply rising (left panels) and flat (right panels) profiles. Note that the accuracy achieved in this
analysis is close to the upper limit, since statistical errors become comparable to the internal dispersion
of the anisotropy profiles in the cluster samples (see e.g. the bottom right panel of Fig.~\ref{ani_prof}).
Still, one may still expect to improve the results and minimize systematic errors by adding more
clusters.

\section{Discussion}

We have performed the Bayesian analysis of mock phase space diagrams of galaxy clusters in terms of a fully
anisotropic model of the phase space density. Our approach allows to constrain parameters of the total mass
profile, $M_{s}$ and $r_{s}$, as well as the asymptotic values of the anisotropy profile, $\beta_{0}$
and $\beta_{\infty}$. The phase space model was designed to detect the radial variation of the
anisotropy profile within a fixed distance range covering two orders of magnitude in radius around $r_{s}$.
The choice of this radial range is motivated by the results from cosmological simulations (Wojtak et al.
2008) and our goal to avoid degeneracy between radial scales and the asymptotes of the anisotropy.

Parameters of the mass profile are determined in our approach with rather satisfying average accuracy of
about 30 percent. On the other hand, around 15 percent of the results are probably subject to systematic error 
and differ from the true values by $50-90$ percent. 
%This effect is probably caused by substructures 
%seen along the line of sight and affecting the scale velocity $V_{s}\propto(M_{s}/r_{s})^{1/2}$ 
%or the scale radius $r_{s}$. 
Nevertheless, it is worth noting that the typical relative
errors of 30 percent in the virial mass are comparable to the potential accuracy of mass determination
from other methods, such as modelling of X-ray gas (e.g. Nagai, Vikhlinin \& Kravtsov 2007), analysis
of velocity moments (Sanchis, {\L}okas \& Mamon 2004) or the standard virial mass estimator (e.g. Biviano
et al. 2006). The constraints on the concentration parameter are however more reliable and tighter in
comparison to the approach based on velocity moments (Sanchis, {\L}okas \& Mamon 2004;
{\L}okas et al. 2006).

We find that the scale mass $M_{s}$
and the virial mass $M_{v}$ tend to be underestimated on average by 11 and 15 percent respectively. It is very
likely that this bias is caused by small velocity interlopers, as described in Biviano et al. (2006), which
are not tractable by algorithms for interloper removal and effectively decrease the velocity dispersion of a
system. Interestingly, a similar offset in the virial mass estimates is noticed in the analysis of mock X-ray
data of clusters (e.g. Ameglio et al. 2009; Nagai et al. 2007). However, this happens
probably incidentally, since these authors claim that their bias is due to the lack of hydrostatic
equilibrium in the outer parts of clusters.

The constraints on the anisotropy profile of a single cluster are barely conclusive. The reason for this
is the presence of substructures at large clustercentric distances which gives rise to significant
systematic errors of $\beta_{\infty}$. The final effect is that, although the
central asymptotic anisotropy is determined rather precisely, the overall constraint on the anisotropy
profile is rather poor. This situation changes in the case of a joint analysis of several phase
space diagrams. We find that it is then easy to obtain reliable constraints on the radial
variation of the anisotropy within the radial range of two orders of magnitude around the scale radius
$r_{s}$. Note that we adopted the phase space scaling that is consistent with the intrinsic parameters of the
mass profile: $M_{s}$ and $r_{s}$, in contrast to commonly used assumption that a sample of phase space
diagrams is homologous with respect to scaling by the virial radius and virial velocity (e.g. van der Marel
et al. 2000; {\L}okas et al. 2006).

In this work we demonstrated the potential and discussed the reliability of the analysis of kinematic data of
galaxy clusters in terms of an anisotropic phase space density model. The method is able to provide robust
constraints on the parameters of the total mass profile and the mean anisotropy profile. The results could
certainly contribute to tests of the mass-concentration relation as well as improve
constraints on the mean anisotropy profile of galaxy clusters (e.g. van der Marel et al. 2000; Biviano \&
Katgert 2004; {\L}okas et al. 2006). This will be the subject of a follow-up work where we analyze real
kinematic data for a sample of $\sim20$ nearby galaxy clusters.

\section*{Acknowledgments}

The simulations have been performed at the Altix of the LRZ Garching.
RW is grateful for the hospitality of Institut d'Astrophysique de Paris and Astrophysikalisches Institut Potsdam
where parts of this work were done. This work was partially supported by the Polish
Ministry of Science and Higher Education under grant NN203025333 as well as
by the Polish-German exchange program of Deutsche Forschungsgemeinschaft
and the Polish-French collaboration program of LEA Astro-PF. RW acknowledges 
support from the START Fellowship for Young Researchers granted by the Foundation 
for Polish Science.

\end{document}